\documentclass[12pt,reqno]{amsart}
\allowdisplaybreaks[4]
\usepackage{graphicx} 
\usepackage{amssymb}
\usepackage{amsmath}
\usepackage{cite}
\def\Z{{\mathbb{Z}}}

\numberwithin{equation}{section}
\makeatletter      
\@addtoreset{equation}{section}
\makeatother       

\begin{document}

\title{String Equations  of
the q-KP Hierarchy}

\author{Kelei Tian\dag, Jingsong He$^*$\dag$\scriptscriptstyle\,$\ddag,   Yucai Su\dag\   and  Yi
Cheng\dag}
\dedicatory { \dag\ Department of Mathematics, University of Science and Technology of China, Hefei, 230026 Anhui,  China\\
\ddag\  Department of Mathematics, Ningbo University, Ningbo, 315211
Zhejiang, China
 }

\thanks{$^*$Corresponding author. Email: jshe@ustc.edu.cn, hejingsong@nbu.edu.cn}

\begin{abstract}
Based on the Lax operator $L$ and Orlov-Shulman's $M$ operator, the
string equations of the $q$-KP hierarchy are established from
special additional symmetry flows, and the negative Virasoro
constraint generators \{$L_{-n}, n\geq1$\} of the $2-$reduced $q$-KP
hierarchy are also obtained.
\end{abstract}


 \maketitle



PACS(2003):\ 02.30.Ik


\section{Introduction}
The $q$-deformed integrable system (also called $q$-analogue or
$q$-deformation  of classical integrable system) is defined by means
of $q$-derivative $\partial_q $  \cite{ks,kac} instead of usual
derivative $\partial $ with respect to~$x$  in a classical system.
It reduces to a classical integrable system as $q \rightarrow1$.
Recently, the $q$-deformed Kadomtsev-Petviashvili ($q$-KP) hierarchy
is a subject of intensive study in the literature
 from \cite{zhang} to \cite{he}. Its infinite conservation laws, bi-Hamiltonian structure, $\tau$
function, additional symmetries and its constrained sub-hierarchy
have already been reported in \cite{wuzhang,mas,iliev2,tu,he}.

The additional symmetries, string equations and Virasoro constraints
of the  KP hierarchy are important as they are involved in the
matrix models of the string theory \cite{Morozov1994}. For example,
there are several new works
\cite{Morozov1998,Alexandrov,Mironov2008,Aratyn2003,tu2} on this topic. The additional
symmetries were discovered independently at least twice by Sato
School \cite{dkjm} and Orlov-Shulman \cite{os}, in quite different
environments and philosophy although they are equivalent
essentially. It is well-known that L.A.Dickey \cite{dickey1}
presented a very elegant and compact proof of Adler-Shiota-van
Moerbeke (ASvM) formula \cite{asv1,asv2} based on the Lax operator
$L$ and Orlov and Shulman's
 $M$ operator \cite{os}, and gave the string equation and the action of
the additional symmetries on the $\tau$ function of the classical KP
hierarchy. S.Panda and S.Roy gave  the Virasoro and $W$-constraints
on the $\tau$ function of the $p$-reduced KP hierarchy by expanding
the additional symmetry operator in terms of the Lax operator
\cite{panda1,panda2}. It is quite interesting to study the analogous
properties of $q$-deformed KP hierarchy by this expanding method .
The main purpose of this article is to give the string equations of
the $q$-KP hierarchy, and then study the negative Virasoro
constraint generators \{$L_{-n}, n\geq1$\} of $2$-reduced $q$-KP
hierarchy.

The organization of this paper is as follows. We recall some basic
results  and  additional symmetries of $q$-KP hierarchy in Section
2. The string equations are given in Sections 3. The Virasoro
constraints on the $\tau$ function of the 2-reduced ($q$-KdV)
hierarchy are studied in Section 4. Section 5 is devoted to
conclusions and discussions. \vspace{8pt}

At the end of the this section, we shall collect  some useful facts
of $q$-calculus \cite{kac} to make this paper be self-contained. The
$q$-derivative $\partial_q$ is defined by
\begin{equation}
   \partial_q(f(x))=\frac{f(qx)-f(x)}{x(q-1)} \label{q-derivative}
   \end{equation}
and the $q$-shift operator is
\begin{equation}
  \theta(f(x))=f(qx).
  \end{equation}
$\partial_q(f(x))$ recovers the ordinary differentiation
$\partial_x(f(x)) $ as $q$ goes to 1.  Let $\partial_q^{-1}$ denote
the formal inverse of $\partial_q$. In general the following
$q$-deformed Leibnitz rule holds
\begin{equation}
     \partial_q^n \circ f=\sum_{k\ge0}\binom{n}{k}_q\theta^{n-k}(\partial_q^kf)\partial_q^{n-k},\hspace{1cm}n\in
     \Z
     \end{equation}
where the $q$-number and the $q$-binomial are defined by
\[
  (n)_q=\frac{q^n-1}{q-1},
  \]
\[
   \binom{n}{k}_q=\frac{(n)_q(n-1)_q\cdots(n-k+1)_q}{(1)_q(2)_q\cdots(k)_q},\hspace{0.5cm}\binom{n}{0}_q=1.
   \]
For a $q$-pseudo-differential operator($q$-PDO) of the form
$P=\sum_{i=-\infty}^np_i\partial_q^i$, we separate $P$ into the
differential part $P_+=\sum_{i\ge0}p_i\partial_q^i$ and the integral
part $P_-=\sum_{i\leq-1}p_i\partial_q^i$. The conjugate operation
``$*$'' for $P$ is defined by $P^*=\sum\limits_i(\partial_q^*)^ip_i$
with
$\partial_q^*=-\partial_q\theta^{-1}=-\frac{1}{q}\partial_{\frac{1}{q}}$,
$(\partial_q^{-1})^*=(\partial_q^*)^{-1}=-\theta
\partial_q^{-1}$.

The $q$-exponent $e_q^x$ is defined as follows
$$
e_q^x=\sum_{n=0}^{\infty}\dfrac{x^n}{(n)_q!}, \quad (n)_q!=(n)_q
(n-1)_q (n-2)_q\cdots (1)_q.
$$
Its equivalent expression is of the form
\begin{equation} \label{qexp}
e_q^x=\exp(\sum_{k=1}^{\infty}\frac{(1-q)^k}{k(1-q^k)}x^k),
\end{equation}
which is crucial to develop the $\tau$  function of the $q$-KP
hierarchy \cite{iliev2}.

\vspace{8pt}

\section{$q$-KP hierarchy and its additional symmetries}
Similar to the general way of describing the classical KP hierarchy
\cite{dkjm,dickey2}, we first give a brief introduction of $q$-KP
hierarchy  and its additional symmetries based on \cite{iliev2,tu}.

Let $L$ be one $q$-PDO given by
\begin{equation}
L=\partial_q+ u_0 +
u_{-1}\partial_q^{-1}+u_{-2}\partial_q^{-2}+\cdots,
\end{equation}
which are called Lax operator of $q$-KP hierarchy. There exist
infinite number of $q$-partial differential equations related to
dynamical variables $\{u_i(x,t_1, t_2, t_3,\cdots,),i=0,-1,-2, -3,
\cdots \}$ and can be deduced from the generalized Lax equation,
\begin{equation}
\dfrac{\partial L}{\partial t_n}=[B_n, L], \ \ n=1, 2, 3, \cdots,
\end{equation}
which are called $q$-KP hierarchy. Here
$B_n=(L^n)_+=\sum\limits_{i=0}^n b_i\partial_q^i$  and
$L^n_-=L^n-L^n_+$.  $L$ in eq.~(2.1) can be generated by dressing
operator $S=1+ \sum_{k=1}^{\infty}s_k \partial_q^{-k}$ in the
following way
\begin{equation}
L=S \circ \partial_q \circ S^{-1}.
\end{equation}
Dressing operator $S$ satisfies Sato equation
\begin{equation}
\dfrac{\partial S}{\partial t_n}=-(L^n)_-S, \quad n=1,2, 3, \cdots.
\end{equation}
The $q$-wave function $w_q(x,t;z)$  and the $q$-adjoint function
$w_q^*(x,t;z)$  are given  by
$$w_q=S e_q^{xz} \exp({\sum  _{i=1}^{\infty}t_iz^i}),$$
$$w_q^*(x,t;z)=(S^*)^{-1}|_{x/q}e_{1/q}^{-xz}\exp(-\sum_{i=1}^\infty t_iz^i),$$
which satisfies following linear $q$-differential equations
$$Lw_q=zw_q,\quad L^*|_{x/q}w_q^*=zw_q^*.$$
Here the notation $P|_{x/t}=\sum_iP_i(x/t)t^i\partial_q^i$ is used
for $P=\sum_ip_i(x)\partial_q^i$.

Furthermore, $w_q(x,t;z)$ and $w_q^*(x,t;z)$ can be expressed by
sole function $\tau_q(x;t)$ \cite{iliev2} as
\begin{gather}\label{qwavefunction}
  w_q=\frac{\tau_q(x;t-[z^{-1}])}{\tau_q(x;\overline{t})}
  e_q ^{xz}\exp\left(\sum_{i=1}^{\infty}t_iz^i\right)=
\frac{e_q^{xz}e^{\xi(t,z)}e^{-\sum_{i=1}^\infty\frac{z^{-i}}{i}\partial_i}\tau_q}{\tau_q},
\\
  w_q^{*}=\frac{\tau_q(x;t+[z^{-1}])}{\tau_q(x;t)}e_{1/q}^{-xz}
  \exp\left(-\sum_{i=1}^{\infty}t_iz^i\right)=
\frac{e_{1/q}^{-xz}e^{-\xi(t,z)}e^{+\sum_{i=1}^\infty\frac{z^{-i}}{i}\partial_i}\tau_q}{\tau_q},
  \nonumber
\end{gather}
where
\[
  [z]=\left(z,\frac{z^2}{2},\frac{z^3}{3},\ldots\right).
\]

The following Lemma shows there exist an essential correspondence
between $q$-KP hierarchy and KP hierarchy.

\vspace{0.5cm}

\noindent \textbf{Lemma 1.~}[11] Let
$L_1=\partial+u_{-1}\partial^{-1}+u_{-2}\partial^{-2}+\cdots$, where
$\partial=\partial / \partial x$, be a solution of the classical KP
hierarchy and $\tau$ be its tau function. Then

$$\tau_q(x,t)=\tau(t+[x]_q)$$
is a tau function of the $q$-KP hierarchy associated with Lax
operator $L$ in  eq.~(2.1), where
$$[x]_q=\big( x, \frac{(1-q)^2}{2(1-q^2)}x^2, \frac{(1-q)^3}{3(1-q^3)}x^3,\cdots,
\frac{(1-q)^i}{i(1-q^i)}x^i,\cdots \big).$$

 \vspace{0.5cm}

Define $\Gamma_q$ and Orlov-Shulman's $M$ operator
\begin{equation}
\Gamma_q=\sum_{i=1}^{\infty}\Big(it_i+\dfrac{(1-q)^i}{(1-q^i)}x^i\Big)\partial
_q^{i-1},
\end{equation}
\begin{equation}
M= S \Gamma_q S^{-1}.
\end{equation}
Dressing $[\partial_k-\partial_q^k,\Gamma_q]$=0 gives
\begin{equation}
\partial _k M=[B_k,M].
\end{equation}
Eq.~(2.2) together with eq.~(2.8) implies that
\begin{equation}
\partial _k (M^mL^n)=[B_k,M^mL^n].
\end{equation}
Define the additional flows for each pair $m,n$ as follows
\begin{equation}
\dfrac{\partial S}{\partial t_{m,n}^*}=-(M^mL^n)_-S,
\end{equation}
or equivalently
\begin{equation}
\dfrac{\partial L}{\partial t_{m,n}^*}=-[(M^mL^n)_-,L],
\end{equation}
\begin{equation}
\dfrac{\partial M}{\partial t_{m,n}^*}=-[(M^mL^n)_-,M].
\end{equation}
The additional flows ${\partial_{mn}^*}= \dfrac{\partial }{\partial
t_{m,n}^*}$  commute with the hierarchy, i.e.
$[\partial_{mn}^*,\partial_k]=0$ but do not commute with each other,
so they are additional symmetries [12].  $(M^mL^n)_-$ serves as the
generator of the additional symmetries along the trajectory
parametrized by $t_{m,n}^*$.

\vspace{8pt}

\section{String equations of the $q$-KP hierarchy}
In this section we shall get string equations for the $q$-KP
hierarchy from  special additional symmetry flows. For this, we need
a lemma.

\vspace{8pt}

\noindent \textbf{Lemma 2.}  The following equation

\begin{equation}
[M,L]=-1
\end{equation}
holds.

\vspace{8pt}

\noindent \textbf{Proof.} Direct calculations show that
$$
[\Gamma_q,\partial_q]=\left[\sum_{i=1}^\infty\Big(it_i+\frac{(1-q)^i}{1-q^i}x^i\Big)\partial_q^{i-1},\partial_q\right]$$

$$
=\sum_{i=1}^\infty\Big[\frac{(1-q)^i}{1-q^i}x^i
\partial_q^{i-1},\partial_q\Big]$$

$$
\hspace{2.0cm}=\sum_{i=1}^\infty\frac{(1-q)^i}{1-q^i}\left
(x^i\partial_q^i-(\partial_q\circ x^i)\partial_q^{i-1}\right )$$

$$
\hspace{3.5cm}=\sum_{i=1}^\infty\frac{(1-q)^i}{1-q^i}\left
(x^i\partial_q^i-((\partial_qx^i)+q^ix^i\partial_q)\partial_q^{i-1}\right
)$$

$$ \hspace{3.9cm}=\sum_{i=1}^\infty\frac{(1-q)^i}{1-q^i}\left
((1-q^i)x^i\partial_q^i-\frac{1-q^i}{1-q}x^{i-1}\partial_q^{i-1}\right
)$$

$$
\hspace{2.5cm}=\sum_{i=1}^\infty((1-q)^ix^i\partial_q^i-(1-q)^{i-1}x^{i-1}\partial_q^{i-1})$$$$
=-1,\hspace{3.6cm}$$ where we have used $[t_i,\partial_q]=0$ in the
second step and $\partial_q\circ
x^i=(\partial_qx^i)+q^ix^i\partial_q$ in the fourth step. Then

$$[M,L]=[S\Gamma_qS^{-1},S\partial_qS^{-1}]=S[\Gamma_q,\partial_q]S^{-1}=-1.$$

\hfill $\square$

\vspace{8pt}

 By virtue of Lemma 2, we have

\vspace{8pt}

 \noindent \textbf{Corollary 1.}
$[M,L]=-1$ implies $[M,L^n]=-nL^{n-1}$. Therefore,
\begin{equation} [ML^{-n+1},L^n]=-n.
\end{equation}

\vspace{8pt}

The action of  additional flows $\partial_{1,-n+1}^* $ on $L^n$ are
 $\partial_{1,-n+1}^* L^n=-[(ML^{-n+1})_-,L^n]$, which can be written as
\begin{equation}
\partial_{1,-n+1}^* L^n=[(ML^{-n+1})_+,L^n]+n.
\end{equation}

The following theorem holds by virtue of eq.(3.3).

\vspace{8pt}

 \noindent \textbf{Theorem 1.} If an operator $L$ does not depend on the parameters $t_n$ and the
additional variables $t_{1,-n+1}^*$, then $L^n$ is a purely
differential operator, and  the string equations of the $q$-KP
hierarchy are given by
\begin{equation}
[L^n,\frac{1}{n}(ML^{-n+1})_+]=1, \ n=2,3,4,\cdots
\end{equation}

In view of the additional symmetries and string equations, we can
get the following corollary, which plays a crucial role in the study
of the constraints on the $\tau$ function of the p-reduced $q$-KP
hierarchy.

\vspace{8pt}

 \noindent \textbf{Corollary 2.} If $L^n$ is a  differential
 operator, and $\partial_{1,-n+1}^* S=0$, then
\begin{equation}
(ML^{-n+1})_-=\frac{n-1}{2}L^{-n}, \ n=2,3,4,\cdots
\end{equation}

 \vspace{8pt}

 \noindent \textbf{Proof.}  Since $[M,L]=-1$, it is not difficult to
 obtain$$[M,L^{-n+1}]=(n-1)L^{-n},$$and
 hence
 \begin{equation}(ML^{-n+1})_--(L^{-n+1}M)_-=(n-1)L^{-n}.\end{equation}
Noticing $[(n-1)L^{-n},L^n]=0$, then

$$\ \ \ [(ML^{-n+1})_--(L^{-n+1}M)_-,L^n]=0,\ \
 \mbox{i.e.},$$$$
 [(ML^{-n+1})_-,L^n]=[(L^{-n+1}M)_-,L^n].$$Thus

 $$\begin{array}{l}\partial^*_{1,-n+1}L^n=-[(L^{-n+1}M)_-,L^n]\\
 \hspace{1.9cm}=-\frac{1}{2}[(ML^{-n+1})_-+(L^{-n+1}M)_-,L^n],\end{array}$$or
 equivalently$$\partial^*_{1,-n+1}S=-\frac{1}{2}(ML^{-n+1}+L^{-n+1}M)_-S.$$
Therefore, it follows from the  equation $\partial^*_{1,-n+1}S=0$
that$$(ML^{-n+1}+L^{-n+1}M)_-=0.$$Combining this with (3.6) finishes
the proof. \hfill $\square$

\vspace{8pt}

\section{Constraints on the $\tau$ function of the $q$-KdV hierarchy}
In this section, we mainly study the associated constraints on
$\tau$-function of the 2-reduced $q$-KP  ($q$-KdV) hierarchy from
string equations eq.~(3.4). To this end, we first define residue
${\rm res\,} L= u_{-1}$ of $L$ given by eq.~(2.1) and state two very
useful lemmas.

\vspace{8pt}

 \noindent \textbf{Lemma 3.}  For $n=1,2,3,\cdots$,
 \begin{equation}
 \mbox{res\,}L^n=\frac{\partial^2\log\tau_q}{\partial t_1\partial
 t_n}.
  \end{equation}
 where $\tau_q$ is the $tau$ function of the $q$-KP hierarchy.

 \vspace{8pt}

 \noindent \textbf{Proof.}  Taking the residue of $\frac{\partial S}{\partial
 t_n}=-(L^n)_-S$, we get
 $$\frac{\partial s_1}{\partial
 t_n}=-\mbox{res}((L^n)_-(1+s_1\partial_q^{-1}+s_2\partial_q^{-2}+\cdots))
 =-\mbox{res}(L^n)_-=-\mbox{res\,}L^n.$$Noting that
 $u_0=s_1-\theta(s_1)=-x(q-1)\partial_q s_1=x(q-1)\partial_q\partial_{t_1}\log\tau_q,\ s_1=-\frac{\partial\log\tau_q}{\partial
 t_1}$ (see \cite{he}), then
 $$\mbox{res}L^n=-\frac{\partial s_1}{\partial t_n}=\frac{\partial^2\log\tau_q}{\partial t_1\partial
 t_n}.$$
\hfill $\square$

\vspace{8pt}

 \noindent \textbf{Lemma 4.} Orlov-Shulman's $M$ operator has the expansion of the form
 \begin{equation}
M=\sum_{i=1}^{\infty}\Big(it_i+\dfrac{(1-q)^i}{(1-q^i)}x^i\Big)L^{i-1}
+ \sum_{i=1}^{\infty} V_{i+1}L^{-i-1},
\end{equation}
where
$$V_{i+1}=
-i \sum_{a_1+2a_2+3a_3+\cdots=i}(-1)^{a_1+a_2+\cdots}
\dfrac{(\partial t_1)^{a_1}}{a_1!}\dfrac {(\dfrac{1}{2}\partial
t_2)^{a_2}}{a_2!}\dfrac{(\dfrac{1}{3}\partial
t_3)^{a_3}}{a_3!}\cdots \log \tau_q.
$$

 \vspace{8pt}

 \noindent \textbf{Proof.}  First, we assert $Mw_q=\frac{\partial w_q}{\partial
 z}$. Indeed,  from the identity
$\partial_q^{i-1}e_q^{xz}=z^{i-1}e_q^{xz}$ we have that
 $$Mw_q=S\Gamma_qS^{-1}Se_q^{xz}e^{\xi(t,z)}=S\left(\sum_{i=1}^\infty\Big(it_i+\frac{(1-q)^i}
 {1-q^i}x^i\Big)z^{i-1}\right)e_q^{xz}e^{\xi(t,z)},$$
where $\xi(t,z)={\sum  _{i=1}^{\infty}t_iz^i}$. On the other hand,
 $$\frac{\partial w_q}{\partial z}=\frac{\partial(Se_q^{xz}e^{\xi(t,z)})}{\partial z}=
 S\Big(\frac{\partial e_q^{xz}}{\partial z}e^{\xi(t,z)}+e_q^{xz}\frac{\partial e^{\xi(t,z)}}{\partial
 z}\Big)$$$$
 =S\left(\sum_{i=1}^\infty\Big(it_i+\frac{(1-q)^i}{1-q^i}x^i\Big)z^{i-1}\right)e_q^{xz}e^{\xi(t,z)}.$$
 Thus  the assertion is verified. Next, by a direct calculation from eq.(\ref{qexp}) and eq.(\ref{qwavefunction}),  we have
\begin{equation}\log w_q=\sum_{k=1}^\infty\frac{(1-q)^k}{k(1-q^k)}(xz)^k+\sum_{n=1}^\infty
t_nz^n
+\sum_{N=0}^\infty\frac{1}{N!}(-\sum_{i=1}^\infty\frac{z^{-i}}{i}\partial_i)^N\log\tau_q-\log\tau_q.
\end{equation}
Let $M=\sum_{n=1}^\infty a_nL^{n-1}+\sum_{n=1}^\infty b_nL^{-n}$.
Then in light of $Lw_q=zw_q$ and the assertion mentioned in above,
we obtain
$$\frac{\partial w_q}{\partial z}=Mw_q=(\sum_{n=1}^\infty
a_nL^{n-1}+\sum_{n=1}^\infty b_nL^{-n})w_q,$$ and hence
\begin{equation}\frac{\partial\log w_q}{\partial z}=\frac{1}{w_q}\frac{\partial w_q}{\partial z}=\sum_{n=1}^\infty a_nz^{n-1}+\sum_{n=1}^\infty
b_nz^{-n}.\end{equation} Thus by comparing the coefficients of $z$
in $\frac{\partial\log w_q}{\partial z} $ given by eq.~(4.3) and
eq.~(4.4), $a_i$ and $b_i$ are determined such that $M$ is obtained
as eq.~(4.2). \hfill $\square$

To be an intuitive glance, the first few  $V_{i+1}$ are given as
follows.
$$ \hspace{-10.3cm} V_2=\frac{\partial\log\tau_q}{\partial
t_1},$$
$$\hspace{-8.3cm} V_3=\frac{\partial\log\tau_q}{\partial
t_2}-\frac{\partial^2\log\tau_q}{\partial t_1 ^2}, $$
$$\hspace{-6.2cm} V_4=(\frac{1}{2}\frac{\partial^3}{\partial
t_1^3}-\frac{3}{2}\frac{\partial^2}{\partial t_1\partial
t_2}+\frac{\partial}{\partial t_3})\log \tau_q,$$
$$\hspace{-4.3cm} V_5=(-\frac{1}{3!}\frac{\partial^4}{\partial
t_1^4}-\frac{1}{2}\frac{\partial^2}{\partial t_2
^2}-\frac{4}{3}\frac{\partial^2}{\partial t_1\partial
t_3}+\frac{\partial}{\partial t_4})\log \tau_q,$$
$$\ V_6=(\frac{1}{4!}\frac{\partial^5}{\partial
t_1^5}-\frac{5}{12}\frac{\partial^4}{\partial t_1 ^3 \partial t_3
}+\frac{5}{6}\frac{\partial^3}{\partial t_1 ^2 \partial t_3 }
-\frac{5}{4}\frac{\partial^2}{\partial t_1\partial
t_4}-\frac{5}{6}\frac{\partial^2}{\partial t_2\partial
t_3}+\frac{\partial}{\partial t_5})\log \tau_q.$$

\vspace{8pt}

Now we consider the 2-reduced $q$-KP hierarchy($q$-KdV hierarchy),
by setting $L^2_-=0$ or setting
 \begin{equation}
 L^2=\partial_q^2+(q-1)xu\partial_q+u.
  \end{equation}
To make the following theorem be a compact form, introduce
\begin{equation}
 L_{-n}=\frac{1}{2}\sum_{\stackrel{\scriptstyle i=2n+1}{i\neq0({\rm mod\,}2)}}^\infty
  i \tilde{t}_i\frac{\partial}{\partial
  \tilde{t}_{i-2n}} +\frac{1}{4}\sum_{k+l=n+1}(2k-1)(2l-1) \tilde{t}_{2k-1}  \tilde{t}_{2k-1}
 \end{equation}
and
 \begin{equation}
 \tilde{t}_i=t_i+\frac{(1-q)^i}{i(1-q^i)}x^i, \ i=1,2,3,\cdots.
  \end{equation}

\vspace{8pt}

 \noindent \textbf{Theorem 2.} If $L^2$ satisfies eq.~(3.4),
  the Virasoro constraints
  imposed on the $\tau$-function of the $q$-KdV hierarchy are
  \begin{equation}
  L_{-n}\tau_q=0,\ \ n=1,2,3,\cdots,
  \end{equation}
and the Virasoro commutation relations
\begin{equation}
[L_{-n},L_{-m}]=(-n+m)L_{-(n+m)}, \  m,n=1,2,3,\cdots
 \end{equation}
hold.
 \vspace{8pt}

 \noindent \textbf{Proof.}  For $n=1,2,3,\cdots,$ we have
 \begin{equation}\mbox{res}(ML^{-2n+1})=\mbox{res}(ML^{-2n+1})_-=\mbox{res}(-\frac{2n+1}{2}L^{-2n})_-=0\end{equation}
 with the help of eq.~(3.5).
 Substituting the expansion of $M$ in eq.~(4.2) into eq.~(4.10), then
$$\sum_{i=1}^\infty\Big(it_i+\frac{(1-q)^i}{1-q^i}x^i\Big)\mbox{res\,}L^{i-2n}+\sum_{i=1}^\infty\mbox{res}(V_{i+1}L^{-i-2n})=0,$$
which implies
\begin{equation}\sum_{\stackrel{\scriptstyle i=2n+1}{i\neq0({\rm mod\,}2)}}^\infty
  \Big(it_i+\frac{(1-q)^i}{1-q^i}x^i\Big)\mbox{res}L^{i-2n}+(2n-1)t_{2n-1}+\frac{(1-q)^{2n-1}}{1-q^{2n-1}}x^{2n-1}=0.\end{equation}
Substituting
$\mbox{res\,}L^{i-2n}=\frac{\partial^2\log\tau_q}{\partial
t_1\partial t_{i-2n}}$ into eq.~(4.11),  then performing an
integration  with respect to $t_1$ and multiplying by
$\frac{\tau_q}{2}$, it becomes
$$ \tilde{L} _{-n}\tau_q=0,\ \ n=1,2,3,\cdots,$$
  where$$\tilde{L}_{-n}=\frac{1}{2}\sum_{\stackrel{\scriptstyle i=2n+1}{i\neq0({\rm mod\,}2)}}^\infty
  \Big(it_i+\frac{(1-q)^i}{1-q^i}x^i\Big)\frac{\partial}{\partial
  t_{i-2n}}+\frac{(1-q)^{2n-1}}{1-q^{2n-1}}\cdot\frac{1}{2}t_1x^{2n-1}$$
  \begin{equation}
 \hspace{-2.7cm} +\,\frac{1}{2}(2n-1)t_1t_{2n-1}+C(t_2,t_3,\cdots;x).
  \end{equation}
The integration constant $C(t_2,t_3,\cdots;x)$ with respect to $t_1$
could be the arbitrary function with the parameters
$(t_2,t_3,\cdots;x )$. What we will do is to determine
$C(t_2,t_3,\cdots;x)$ such that $\tilde{L}_{-n}$ satisfy Virasoro
commutation relations.

Let
$$\tilde{t}_i=t_i+\frac{(1-q)^i}{i(1-q^i)}x^i, \ i=1,2,3,\cdots,$$
and choose $C(t_2,t_3,\cdots;x)$ as
 $$C(t_2,t_3,\cdots;x)=- \frac{1}{4} \sum_{k=3}^{2n-3}(2k-1)(2n-2k+1)
 \Big(t_{2k-1}+\frac{(1-q)^{2k-1}}{(2k-1)(1-q^{2k-1})}x^{2k-1}\Big)$$
 $$\hspace{0.8cm}
 \cdot\Big(t_{2n-2k+1}+\frac{(1-q)^{2n-2k+1}}{(2n-2k+1)(1-q^{2n-2k+1})}x^{2n-2k+1}\Big)$$
$$
-\frac{1}{2}(2n-1)x\Big(t_{2n-1}+\frac{(1-q)^{2n-1}}{(2n-1)(1-q^{2n-1})}x^{2n-1}\Big),$$
Then
$$
 \tilde{L}_{-n}=\frac{1}{2}\sum_{\stackrel{\scriptstyle i=2n+1}{i\neq0({\rm mod\,}2)}}^\infty
  i\tilde{t}_i\frac{\partial}{\partial
  \tilde{t}_{i-2n}} +\frac{1}{4}\sum_{k+l=n+1}(2k-1)(2l-1) \tilde{t}_{2k-1}
  \tilde{t}_{2k-1}\equiv L_{-n}
$$
and
$$L_{-n}\tau_q=0,\ \ n=1,2,3,\cdots$$ as we expected. By a straightforward and tedious calculation,  the Virasoro
commutation relations
$$
[L_{-n},L_{-m}]=(-n+m)L_{-(n+m)}, m,n=1,2,3,\cdots$$ can be
verified. \hfill $\square$

 \vspace{12pt}


 \noindent \textbf{Remark 1.} As we know, the $q$-deformed KP hierarchy
 reduces to the classical KP hierarchy when $q \rightarrow 1$ and
 $u_0=0$. The parameters
 $(\tilde{t}_1,\tilde{t}_2,\cdots,\tilde{t}_i,\cdots)$ in eq.~(4.6) tend to
 $(t_1+x,t_2,\cdots,t_i,\cdots)$ as $q \rightarrow 1$. One can
 further
 identify $t_1+x$ with $x$ in the classical KP hierarchy, i.e. $t_1+x \rightarrow
 x$, therefore  the Virasoro generators $L_{-n}$  in eq.~(4.6) of the
2-reduced $q$-KP hierarchy tend to
 \begin{equation}
 \hat{L}_{-n}=\frac{1}{2}\sum_{\stackrel{\scriptstyle i=2n+1}{i\neq0({\rm mod\,}2)}}^\infty
  it_i\frac{\partial}{\partial
 t_{i-2n}} +\frac{1}{4}\sum_{k+l=n+1}(2k-1)(2l-1) t_{2k-1} t_{2k-1},
 n=2,3,\cdots
 \end{equation}
and
  \begin{equation}
 \hat{L}_{-1}=\frac{1}{2}\sum_{\stackrel{\scriptstyle i=3}{i\neq0({\rm mod\,}2)}}^\infty
  it_i\frac{\partial}{\partial
 t_{i-2}} +\frac{1}{4}x^2,
 \end{equation}
 which are identical with the results of the classical KP hierarchy given by L.A.Dickey \cite{dickey3} and S.Panda,
 S.Roy \cite{panda1}.

\section{Conclusions and discussions}
To summarize, we have derived the string equations in eq.~(3.4) and
the negative
 Virasoro constraint generators on the \ $\tau$ function of 2-reduced $q$-KP
 hierarchy in eq.~(4.8)
in Theorem 2. The results  of this paper show obviously that the
Virasoro generators $\{L_{-n},n\geq1\}$ of the $q$-KP hierarchy are
different with the $\{\hat{L}_{-n}, n\geq1\}$ of the KP hierarchy,
although they satisfy the common Virasoro commutation relations.
Furthermore, one can find the following interesting relation between
the $q$-KP hierarchy and the KP hierarchy
$$L_{-n}=\hat{L}_{-n}|_{t_i \rightarrow \tilde{t}_i=t_i+\frac{(1-q)^i}{i(1-q^i)}x^i},$$
and it seems to demonstrate that $q$-deformation  is a non-uniform
transformation for  coordinates $t_i \rightarrow \tilde{t}_i,$ which
is consistent with results on $\tau$ function \cite{iliev2} and the
$q$-soliton \cite{he} of the $q$-KP hierarchy.

\vspace{12pt}
 For the p-reduced $(p\geq 3)$ $q$-KP hierarchy, which is the $q$-KP
hierarchy satisfying the reduction condition $(L^p)_-=0$, we can
obtain   $(ML^{pn+1})_-=0$. Using the similar technique in $q$-KdV
hierarchy, we can deduce the  Virasoro constraints on the $\tau$
function of the p-reduced $q$-KP hierarchy for $p\geq3$. Moreover,
 for $\{L_{n}, n\geq0\}$ we find a subtle point
 at the calculation  of $res(V_{i+1}L^{-i+2n})$, and will try to
 study it in the future.

\vspace{12pt}

{\bf Acknowledgments} {\small This work is supported by the NSF of
China under Grant No. 10671187, 10825101. }



\end{document}